\documentclass[aip,apl,reprint,amsmath,amssymb]{revtex4-1}
\usepackage{graphicx}
\usepackage{dcolumn}
\usepackage{bm}
\usepackage[english]{babel}
\usepackage{epsfig}
\usepackage[cp1251]{inputenc}
\usepackage[mathlines]{lineno}
\usepackage{amsmath,amssymb,graphicx,color}

\begin{document}

\title{Ultracompact all-dielectric superdirective antennas}

\author{\firstname{Alexander ~E.} \surname{Krasnok}}
\affiliation{National Research University of Information
Technologies, Mechanics and Optics (ITMO), \\ St. Petersburg 197101,
Russia}
\author{\firstname{Dmitry ~S.} \surname{Filonov}}
\affiliation{National Research University of Information
Technologies, Mechanics and Optics (ITMO), \\ St. Petersburg 197101,
Russia}
\author{\firstname{Pavel ~A.} \surname{Belov}}
\affiliation{National Research University of Information
Technologies, Mechanics and Optics (ITMO), \\ St. Petersburg 197101,
Russia}
\author{\firstname{Alexey ~P.} \surname{Slobozhanyuk}}
\affiliation{National Research University of Information
Technologies, Mechanics and Optics (ITMO), \\ St. Petersburg 197101,
Russia}
\author{\firstname{Constantin ~R.} \surname{Simovski}}
\affiliation{National Research University of Information
Technologies, Mechanics and Optics (ITMO), \\ St. Petersburg 197101,
Russia} \affiliation{Aalto University, School of Electric and
Electronic Engineering, Aalto FI76000, Finland}
\author{\firstname{Yuri ~S.} \surname{Kivshar}}
\affiliation{National Research University of Information
Technologies, Mechanics and Optics (ITMO), \\ St. Petersburg 197101,
Russia} \affiliation{Nonlinear Physics Center, Research School of
Physics and Engineering, Australian National University, Canberra
ACT 0200, Australia}
\date{\today}

\begin{abstract}
We demonstrate a simple way to achieve superdirectivity of
electrically small antennas based on a spherical dielectric particle
with a notch. We predict this effect theoretically for nanoantennas
excited by a point-like emitter located in the notch, and then
confirm it experimentally at microwaves for a ceramic sphere excited
by a small wire dipole. We explain the effect of superdirectivity by
the resonant excitation of high-order multipole modes of electric
and magnetic fields which are usually negligible for small perfect
spherical particles.
\end{abstract}

\pacs{} 

\maketitle

\begin{figure}[b]
\centerline{\includegraphics[width=9cm]{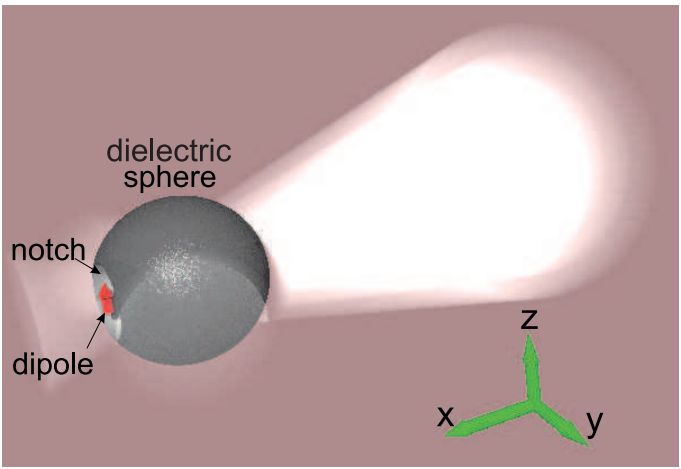}} \caption{
Geometry of the notched all-dielectric nanoantenna.}\label{geometr}
\end{figure}

Electrically small radiating systems whose directivity exceeds
significantly that of a dipole are usually called
superdirective~\cite{Balanis}. Superdirectivity is an important
property of radio-frequency antennas employed for space
communications and radioastronomy, and it can be achieved in antenna
arrays in a narrow frequency range and for a sophisticated system of
phase shifters~\cite{Balanis}. Achieving high radiation directivity
is also important for actively studied optical
nanoantennas~\cite{Hulst_08_OSA,He,Jain,AluWireless,
Novotny_10_NatPhot,Nelson_boock_11,Rodriguez,Hecht_obz_12}. Similar
to radio-frequency antennas, a nanoantenna converts localized
electromagnetic field to freely propagating light, and vice
versa~\cite{Novotny_10_NatPhot, Hecht_obz_12}. For prospective
optical wireless circuits on a chip, nanoantennas are required to be
highly directive and compact~\cite{AluWireless}. In nanophotonics
higher directivity can be achieved in arrayed plasmonic antennas
utilizing the Yagi-Uda design~\cite{Novotny_10_NatPhot,
Hecht_obz_12, Hulst_08_OSA}. However, such nanoantennas have the
size larger than the radiation wavelength $\lambda$, and individual
elements of these arrays are not optically small. Usually, small
plasmonic nanoantennas possess weak directivity close to that of a
point dipole~\cite{Jain, He,Rodriguez}. Despite its importance,
antenna's superdirectivity  in the optical frequency range was not
discussed or demonstrated so far. Recently, it was suggested to use
different dielectric and semiconductor materials for the development
of antennas at the nanoscale~\cite{Krasnok_11, Krasnok_APL_12,
KrasnokOE}. Such all-dielectric nanoantennas consist of
high-permittivity nanoparticles having both resonant electric and
magnetic optical responses~\cite{Krasnok_11,
Krasnok_APL_12,KrasnokOE, Evlyukhin, Kuznetsov}. This approach
allows to study an optical analogue of the so-called Huygens source,
an elementary emitting system with properly balanced electric and
magnetic dipoles oscillating with the same
phase~\cite{Balanis,Krasnok_11, Krasnok_APL_12, KrasnokOE}. As a
result, nearly twice higher directivity than that of a single
electric dipole has been reported. However, this directivity is
still insufficient for nanophotonics applications. Several recent
studies suggested to enhance directivity of nanoantennas by
adjusting the distance between a point-like emitter and plasmonic
nanoparticles~\cite{RollyOL} or employing a core-shell plasmonic
resonator with metamaterials~\cite{Alu}.

In this Letter we reveal, for the first time to our knowledge, a
novel way for achieving superdirectivity of antennas with a
subwavelength (maximum size 0.4-0.5 $\lambda$) radiating system
comprising a point-like emitter and a dielectric particle with a
small notch. This effect is achieved without using complex antenna
arrays, and it is valid for a wide range of frequencies.

First, we demonstrate possibility to create a superdirective
optically small nanoantenna that does not require metamaterial. We
consider one semiconductor nanoparticle with the permittivity $\rm
Re{\varepsilon}=15-16$ radiated by light at wavelength $\lambda$
(for $\lambda=$440-460~nm this corresponds to a nanoparticle made of
crystalline silicon~\cite{Palik}) and the radius $R_{\mbox{s}}=90$
nm being almost five times smaller than $\lambda$. For a perfect
sphere, lower-order multipoles for both electric and magnetic fields
are excited while the contribution of higher-order modes is
negligible~\cite{Krasnok_11, Krasnok_APL_12, KrasnokOE}. However,
making a small notch in the spherical particle breaks the symmetry
allowing the excitation of higher-order multipole moments of the
sphere. This is achieved by placing a nanoemitter (e.g. a quantum
dot) within a small notch created on the sphere surface, as shown in
Fig.~\ref{geometr}. The notch in our example has the shape of a
hemisphere with a radius $R_{\mbox{n}}\ll R_{\mbox{s}}$. The emitter
can be modeled as a point-like dipole and it is shown in the figure
by a red arrow. It turns out that such a small modification of the
sphere would allow the efficient excitation of higher-order
spherical multipole modes.

It is important to mention that our approach is seemingly close to
the idea of Refs.~\cite{Sveta1, Wang} where a small notch on a
surface of a semiconductor microlaser was used to achieve higher
emission directivity by modifying the field distribution inside the
resonator~\cite{Scully}. An important difference between those
earlier studies and our work is that the design discussed earlier is
not optically small and the directive emission is not related to
superdirectivity. In our case, the nanoparticle is much smaller than
the wavelength, and our design allows superdirectivity.

\begin{figure}[t]
\centerline{\includegraphics[width=8cm]{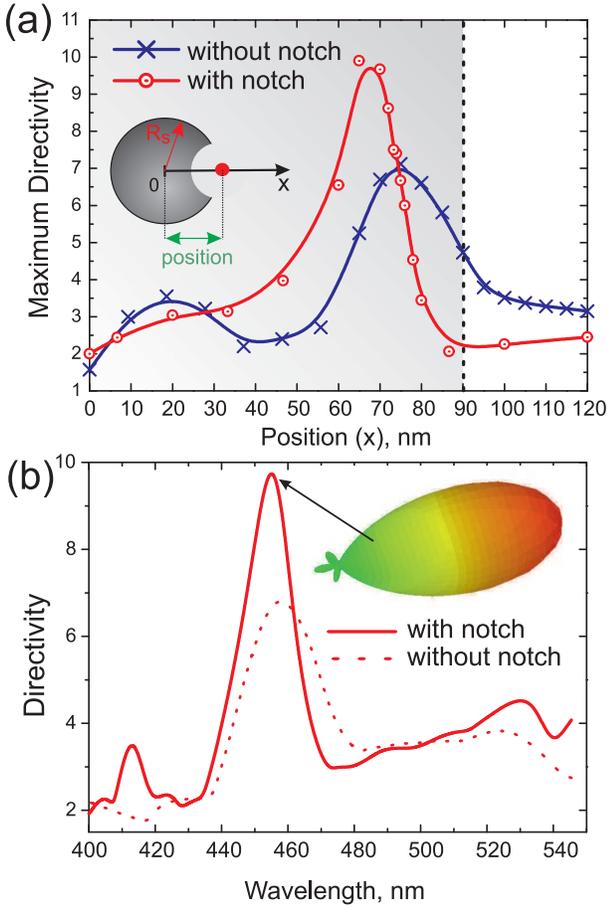}}
\caption{(Color online) (a) Maximum of directivity depending on the
position of the dipole ($\lambda=455$ nm) in the case of a sphere
with and without notch, respectively. Vertical dashed line marks the
particle radius centered at the coordinate system. (b) Directivity
dependence on the radiation wavelength. The inset shows
three-dimensional radiation pattern of the structure
($R_{\mbox{s}}=90$~nm and $R_{\mbox{n}}=40$~nm).} \label{direct}
\end{figure}

To study the problem numerically, we employ the software package CST
Microwave Studio with the super-computing system Tesla S2050.
Figure~\ref{direct}(a) shows the dependence of the maximum
directivity $D_{\mbox{mav}}$ on the position of the emitting dipole
in the case of a sphere $R_{\mbox{s}}=90$~nm without a notch, at the
wavelength $\lambda=455$~nm (blue curve with crosses). This
dependence has the maximum ($D_{\mbox{max}}=7.1$) when the emitter
is placed inside the particle at the distance 20~nm from its
surface. The analysis shows that in this case the electric field
distribution inside a particle corresponds to the noticeable
excitation of higher-order multipole modes. This becomes possible
due to strong inhomogeneity of the external field produced by the
nanoemitter. Furthermore, the excitation of higher-order multipoles
can be significantly improved by making a small notch in the silicon
spherical nanoparticle and placing the emitter inside that notch, as
shown in Fig.~\ref{geometr}. This modification of the nanoparticle
transforms it into a resonator for high-order multipole moments.

\begin{figure}[t]
\centerline{\includegraphics[width=9cm]{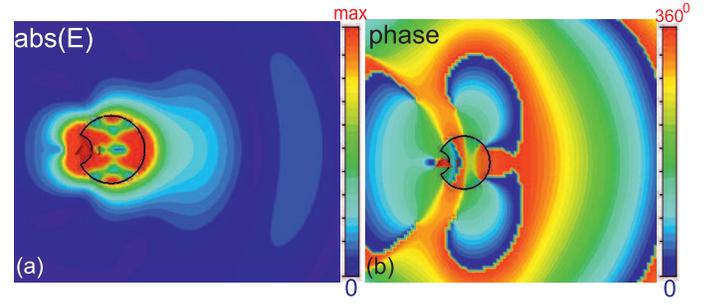}}
\caption{Distribution of (a) absolute values and (b) phases of the
electric field at the wavelength $\lambda=455$ nm.}
\label{fieldfree}
\end{figure}

In our problem, the notch has the form of a hemisphere with the
center it the dielectric nanoparticle's surface. The optimal radius
of the notch is $R_{\mbox{n}}=40$ nm, that we find by means of
numerical optimization. Red curve with circles in the
Fig.~\ref{direct}(a) shows maximum of directivity corresponding to
this geometry. Maximal directivity at wavelength 455~nm is
$D_{\mbox{max}}=10$. Figure~\ref{direct}(b) shows the dependence of
directivity on the wavelength with and without a notch. The inset
shows the three-dimensional radiation pattern of the structure at
$\lambda=$455 nm. This pattern has an angular width (at the power of
3 dB) of the main lobe of $40^{\circ}$.

Expansion of the field into multipoles offers an illustrative
description of the internal field composition for the spherical
particle. This is given by a series of spherical harmonics with the
coefficients $a_{E}(l,m)$ and $a_{M}(l,m)$, which characterize the
electrical and magnetic multipole moments~\cite{Jackson},
\begin{eqnarray}
a_{E}(l,m)&=&-\frac{4\pi i
k^{l+2}}{(2l+1)!!}\left(\frac{l+1}{l}\right)^{1/2}Q_{lm},\label{qe1}
\\ a_{M}(l,m)&=&\frac{4\pi
ik^{l+2}}{(l+1)(2l+1)!!}\left(\frac{l+1}{l}\right)^{1/2}M_{lm},
\label{qe2}
\end{eqnarray}
where
\begin{eqnarray}
Q_{lm}=\int_{V}r^lY^*_{lm}\rho d^3x,~~
M_{lm}=\int_{V}r^lY^*_{lm}\mbox{div}\left(\frac{\mathbf{j}\times\mathbf{r}}{c}\right)d^3x,\nonumber
\end{eqnarray}
$\rho$ and $\mathbf{j}$ are densities of the induced electrical
charges and polarization currents that can be easily expressed
through the internal field and the complex permittivity of the
sphere, $Y_{lm}$ - spherical harmonics of the orders $(l>0,0\ge m\le
l)$, and $k=2\pi/\lambda$, $c$ is the speed of light. Coefficients
$a_{E}(l,m)$ and $a_{M}(l,m)$ determine the electric and magnetic
mutipole moments (namely, dipole at $l=1$, quadrupole at $l=2$,
etc.). In the coordinate system shown in Fig.~1(a), we can write the
electric and magnetic dipole moments ($\bf p$ and $\bf m$,
respectively) in the form~\cite{Jackson},
\begin{eqnarray}
\begin{bmatrix}
a_E(1,0)\\ a_M(1,0)
\end{bmatrix}=\sqrt{3\over 4\pi}
\begin{bmatrix}
p_z\\ \sqrt{\varepsilon_0/ \mu_0} m_z
\end{bmatrix},\quad\nonumber\\
\begin{bmatrix}
a_E(1,1)\\ a_M(1,1)
\end{bmatrix}=\sqrt{3\over 8\pi}
\begin{bmatrix}
(p_x-i p_y)\\ \sqrt{\varepsilon_0/\mu_0}(m_x-i m_y)
\end{bmatrix}.\nonumber
\end{eqnarray}

In general, the multipole coefficients determine not only the mode
structure of the internal field but also the angular distribution of
the radiation. Internal field was calculated numerically with the
expansion into the multipole series. Figures~\ref{fieldfree}(a,b)
show the distribution of the absolute values and phases of the
electric field. Electric (as well as magnetic) field intensity
inside the particle is strongly inhomogeneous at $\lambda=$455~nm
(i.e. in the regime of maximal directivity). In this regime, the
internal area where the electric field oscillates with approximately
the same phase is maximal. This area is located near the back side
of the spherical particle, as can be seen in
Fig.~\ref{fieldfree}(b). In other words, in the regime of maximal
directivity the nanoparticle acts as a device which straightens the
surfaces with same phase of the emitter's near field.
\begin{figure}[t]
\centerline{\includegraphics[width=8cm]{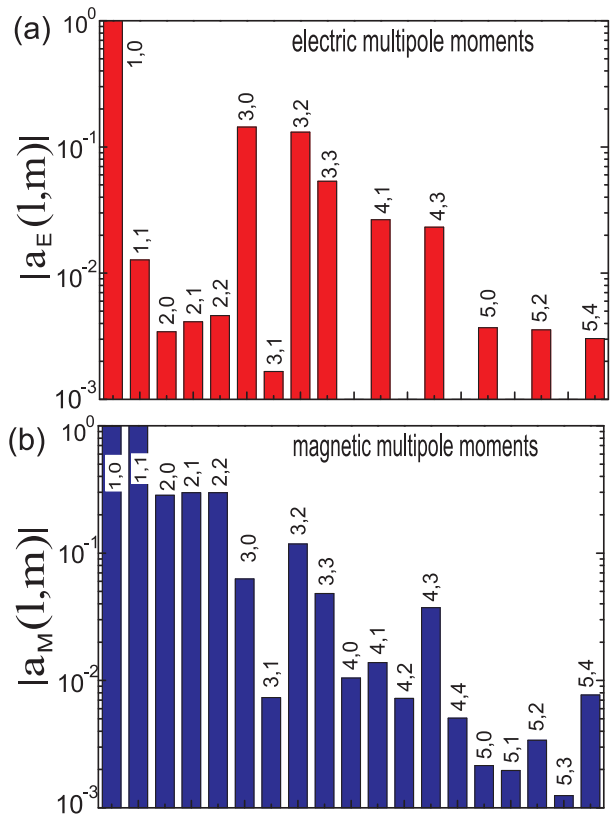}}
\caption{Absolute values of (a) electric and (b) magnetic multipole
moments that provide the main contribution to the radiation of
antenna at the wavelength 455~nm. Particle and notch radii are equal
to $R_{\mbox{s}}=90$ nm and $R_{\mbox{n}}=40$ nm, respectively.}
\label{moments}
\end{figure}

\begin{figure}[t]
\centerline{\includegraphics[width=9cm]{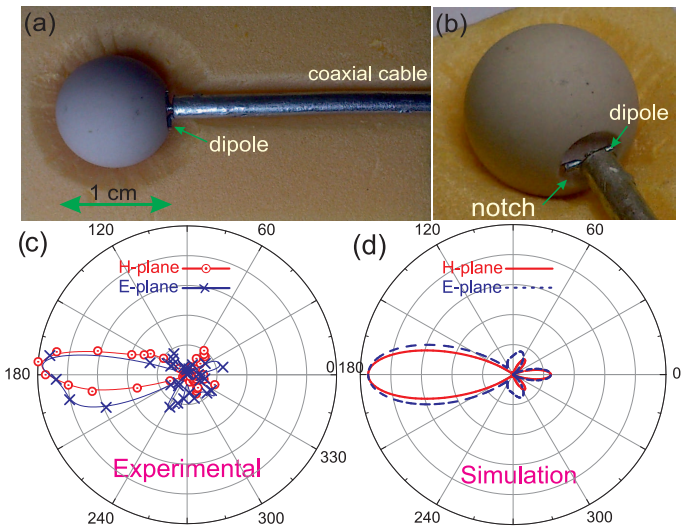}}
\caption{Photographs of (a) top view and (b) perspective view of a
notched all-dielectric microwave antenna. Experimental (c) and
numerical (d) radiation patterns of the antenna in both $E$- and
$H$-planes at the frequency 16.8 GHz. The crosses and circles
correspond to the experimental data.} \label{experiment}
\end{figure}

The values of high-order multipole moments for this electric field
distribution are calculated with the use of Eqs.~(\ref{qe1}) and
(\ref{qe2}), and they are shown in Figs.~\ref{moments}(a,b), where
we observe strong high-order multipoles excited together with the
electric and magnetic dipoles $a_{E}(1,1)$, $a_{E}(1,0)$,
$a_{M}(1,1)$, and $a_{M}(1,0)$. We notice that the absolute values
of all magnetic moments are larger than those of the electric
moments in the corresponding multipole orders. The spectrum of
noticeably excited magnetic moments is also larger than that for the
electric moments. Physically, this means that the presence of the
notch has the greatest impact on the value of the polarization
current $\mathbf{j}$, whereas the distribution of the induced
charges $\rho$ is weakly affected by the notch.

We have confirmed the generality of the predicted effect by studying
the similar problem in the microwave range. To do this, we scale up
the nanoantenna discussed above to the microwave frequencies. As a
high-permittivity all-dielectric antenna in this frequency range, we
employ MgO-TiO$_{2}$ ceramic ~\cite{Krasnok_APL_12} characterized at
microwaves by dielectric constant of 16 and dielectric loss factor
of (1.12-1.17)10$^{-4}$. We use the particle with the radius of
$R_{\mbox{s}}=5$~mm and apply a small vibrator~\cite{Balanis}
excited by a coaxial cable [see Figs.~\ref{experiment}(a,b)]. The
size of the hemispherical notch is approximately equal to
$R_{\mbox{n}}=2$ mm. Styrofoam material with the dielectric
permittivity close to 1 is used to fix the antenna in the
azimuthal-rotation unit, as shown in Figs.~\ref{experiment}(a,b).
First, we perform numerical simulations of this antenna in CST
Microwave Studio, and observe high directivity of this dielectric
antenna at the frequency 16.8 GHz. Next, we study experimentally the
radiation pattern of the antenna in the anechoic chamber. The
results of the experimental measurements and numerical simulation of
the radiation pattern in both $E$- and $H$-planes are summarized in
Figs.~\ref{experiment}(c),(d). Radiation patterns in both the planes
are narrow beams with a lobe angle about 35$^{\circ}$. Experimental
coefficients of the directivity in both $E$- and $H$-planes are
equal to 5.9 and 8.4, respectively. The numerical values of these
quantities are equal to 6.8 and 8.1. Our experimental data is in a
very good agreement with the numerical results. The small difference
between the experimental and numerical results in the E plane, can
be explained by the asymmetry of the dipole.

Note, that the observed directivity is the same as one of
all-dielectric Yagi-Uda antenna with size about 2
wavelength~\cite{Krasnok_APL_12}. However, the total size of our
antenna with notch is $\sim\lambda/2.5$. Thus, our experiment is
clearly demonstrating superdirectivity performance.

In conclusion, we have proposed a novel and frequency-independent
approach to achieve antenna's superdirectivity through the
excitation of electric and magnetic higher-order multipole modes in
the electrically small dielectric particle with a noth. In the
visible frequency range, we have studied this effect numerically for
a subwavelength spherical c-Si nanoparticle with a notch on its
surface. We have demonstrated that by locating a nanoemitter in the
notch provides a strong excitation of higher-order modes and
therefore high values of directivity not achievable for small
spheres without a notch. Then, we have demonstrated experimentally
that microwave antennas with a notch demonstrate the similar
property of superdirectivity. Our results introduce an important
concept of all-dielectric notched antennas, which is applicable to a
wide frequency range and would allow creating a new generation of
highly directed emitters with a variety of useful applications.

The authors thank P. V. Kapitanova for valuable discussions. This
work was supported by the Ministry of Education and Science of the
Russian Federation, Dynasty Foundation (Russia), and the Australian
Research Council.


\begin{thebibliography}{21}%
\makeatletter
\providecommand \@ifxundefined [1]{%
 \@ifx{#1\undefined}
}%
\providecommand \@ifnum [1]{%
 \ifnum #1\expandafter \@firstoftwo
 \else \expandafter \@secondoftwo
 \fi
}%
\providecommand \@ifx [1]{%
 \ifx #1\expandafter \@firstoftwo
 \else \expandafter \@secondoftwo
 \fi
}%
\providecommand \natexlab [1]{#1}%
\providecommand \enquote  [1]{``#1''}%
\providecommand \bibnamefont  [1]{#1}%
\providecommand \bibfnamefont [1]{#1}%
\providecommand \citenamefont [1]{#1}%
\providecommand \href@noop [0]{\@secondoftwo}%
\providecommand \href [0]{\begingroup \@sanitize@url \@href}%
\providecommand \@href[1]{\@@startlink{#1}\@@href}%
\providecommand \@@href[1]{\endgroup#1\@@endlink}%
\providecommand \@sanitize@url [0]{\catcode `\\12\catcode
`\$12\catcode
  `\&12\catcode `\#12\catcode `\^12\catcode `\_12\catcode `\%12\relax}%
\providecommand \@@startlink[1]{}%
\providecommand \@@endlink[0]{}%
\providecommand \url  [0]{\begingroup\@sanitize@url \@url }%
\providecommand \@url [1]{\endgroup\@href {#1}{\urlprefix }}%
\providecommand \urlprefix  [0]{URL }%
\providecommand \Eprint [0]{\href }%
\providecommand \doibase [0]{http://dx.doi.org/}%
\providecommand \selectlanguage [0]{\@gobble}%
\providecommand \bibinfo  [0]{\@secondoftwo}%
\providecommand \bibfield  [0]{\@secondoftwo}%
\providecommand \translation [1]{[#1]}%
\providecommand \BibitemOpen [0]{}%
\providecommand \bibitemStop [0]{}%
\providecommand \bibitemNoStop [0]{.\EOS\space}%
\providecommand \EOS [0]{\spacefactor3000\relax}%
\providecommand \BibitemShut  [1]{\csname bibitem#1\endcsname}%
\let\auto@bib@innerbib\@empty
\bibitem [{\citenamefont {Balanis}(1982)}]{Balanis}%
  \BibitemOpen
  \bibfield  {author} {\bibinfo {author} {\bibfnamefont {C.}~\bibnamefont
  {Balanis}},\ }\href@noop {} {\emph {\bibinfo {title} {Antenna Theory:
  Analysis and Design}}}\ (\bibinfo  {publisher} {New York ; Wiley},\ \bibinfo
  {year} {1982})\BibitemShut {NoStop}%
\bibitem [{\citenamefont {Taminiau}, \citenamefont {Stefani},\ and\
  \citenamefont {Hulst}(2008)}]{Hulst_08_OSA}%
  \BibitemOpen
  \bibfield  {author} {\bibinfo {author} {\bibfnamefont {T.~H.}\ \bibnamefont
  {Taminiau}}, \bibinfo {author} {\bibfnamefont {F.~D.}\ \bibnamefont
  {Stefani}}, \ and\ \bibinfo {author} {\bibfnamefont {N.~F.}\ \bibnamefont
  {Hulst}},\ }\href@noop {} {\bibfield  {journal} {\bibinfo  {journal} {Optics
  Express}\ }\textbf {\bibinfo {volume} {16}},\ \bibinfo {pages} {10858}
  (\bibinfo {year} {2008})}\BibitemShut {NoStop}%
\bibitem [{\citenamefont {He}\ \emph {et~al.}(2009)\citenamefont {He},
  \citenamefont {Cui}, \citenamefont {Ye}, \citenamefont {Zhang},\ and\
  \citenamefont {Jin}}]{He}%
  \BibitemOpen
  \bibfield  {author} {\bibinfo {author} {\bibfnamefont {S.}~\bibnamefont
  {He}}, \bibinfo {author} {\bibfnamefont {Y.}~\bibnamefont {Cui}}, \bibinfo
  {author} {\bibfnamefont {Y.}~\bibnamefont {Ye}}, \bibinfo {author}
  {\bibfnamefont {P.}~\bibnamefont {Zhang}}, \ and\ \bibinfo {author}
  {\bibfnamefont {Y.}~\bibnamefont {Jin}},\ }\href@noop {} {\bibfield
  {journal} {\bibinfo  {journal} {Materials Today}\ }\textbf {\bibinfo {volume}
  {12}},\ \bibinfo {pages} {16} (\bibinfo {year} {2009})}\BibitemShut {NoStop}%
\bibitem [{\citenamefont {Jain}\ and\ \citenamefont {El-Sayed}(2010)}]{Jain}%
  \BibitemOpen
  \bibfield  {author} {\bibinfo {author} {\bibfnamefont {P.~K.}\ \bibnamefont
  {Jain}}\ and\ \bibinfo {author} {\bibfnamefont {M.~A.}\ \bibnamefont
  {El-Sayed}},\ }\href@noop {} {\bibfield  {journal} {\bibinfo  {journal}
  {Chemical Physics Letters}\ }\textbf {\bibinfo {volume} {487}},\ \bibinfo
  {pages} {153–164} (\bibinfo {year} {2010})}\BibitemShut {NoStop}%
\bibitem [{\citenamefont {Alu}\ and\ \citenamefont
  {Engheta}(2010)}]{AluWireless}%
  \BibitemOpen
  \bibfield  {author} {\bibinfo {author} {\bibfnamefont {A.}~\bibnamefont
  {Alu}}\ and\ \bibinfo {author} {\bibfnamefont {N.}~\bibnamefont {Engheta}},\
  }\href@noop {} {\bibfield  {journal} {\bibinfo  {journal} {Phys. Rev. Lett.}\
  }\textbf {\bibinfo {volume} {104}},\ \bibinfo {pages} {213902} (\bibinfo
  {year} {2010})}\BibitemShut {NoStop}%
\bibitem [{\citenamefont {Novotny}\ and\ \citenamefont {van
  Hulst}(2011)}]{Novotny_10_NatPhot}%
  \BibitemOpen
  \bibfield  {author} {\bibinfo {author} {\bibfnamefont {L.}~\bibnamefont
  {Novotny}}\ and\ \bibinfo {author} {\bibfnamefont {N.}~\bibnamefont {van
  Hulst}},\ }\href@noop {} {\bibfield  {journal} {\bibinfo  {journal} {Nat.
  Photon.}\ }\textbf {\bibinfo {volume} {5}},\ \bibinfo {pages} {83} (\bibinfo
  {year} {2011})}\BibitemShut {NoStop}%
\bibitem [{\citenamefont {Mahendra}\ and\ \citenamefont
  {Nelson}(2011)}]{Nelson_boock_11}%
  \BibitemOpen
  \bibfield  {author} {\bibinfo {author} {\bibfnamefont {R.}~\bibnamefont
  {Mahendra}}\ and\ \bibinfo {author} {\bibfnamefont {D.}~\bibnamefont
  {Nelson}},\ }\href@noop {} {\emph {\bibinfo {title} {Metal Nanoparticles in
  Microbiology}}}\ (\bibinfo  {publisher} {Springer},\ \bibinfo {year}
  {2011})\BibitemShut {NoStop}%
\bibitem [{\citenamefont {Rodriguez}\ \emph {et~al.}(2012)\citenamefont
  {Rodriguez}, \citenamefont {Murai}, \citenamefont {Verschuuren},\ and\
  \citenamefont {Rivas}}]{Rodriguez}%
  \BibitemOpen
  \bibfield  {author} {\bibinfo {author} {\bibfnamefont {S.~R.~K.}\
  \bibnamefont {Rodriguez}}, \bibinfo {author} {\bibfnamefont {S.}~\bibnamefont
  {Murai}}, \bibinfo {author} {\bibfnamefont {M.~A.}\ \bibnamefont
  {Verschuuren}}, \ and\ \bibinfo {author} {\bibfnamefont {J.~G.}\ \bibnamefont
  {Rivas}},\ }\href@noop {} {\bibfield  {journal} {\bibinfo  {journal} {Phys.
  Rev. Lett.}\ }\textbf {\bibinfo {volume} {109}},\ \bibinfo {pages} {166803}
  (\bibinfo {year} {2012})}\BibitemShut {NoStop}%
\bibitem [{\citenamefont {Biagioni}, \citenamefont {Huang},\ and\ \citenamefont
  {Hecht}(2012)}]{Hecht_obz_12}%
  \BibitemOpen
  \bibfield  {author} {\bibinfo {author} {\bibfnamefont {P.}~\bibnamefont
  {Biagioni}}, \bibinfo {author} {\bibfnamefont {J.}~\bibnamefont {Huang}}, \
  and\ \bibinfo {author} {\bibfnamefont {B.}~\bibnamefont {Hecht}},\
  }\href@noop {} {\bibfield  {journal} {\bibinfo  {journal} {Rep. Prog. Phys.}\
  }\textbf {\bibinfo {volume} {75}},\ \bibinfo {pages} {024402} (\bibinfo
  {year} {2012})}\BibitemShut {NoStop}%
\bibitem [{\citenamefont {Krasnok}\ \emph {et~al.}(2011)\citenamefont
  {Krasnok}, \citenamefont {Miroshnichenko}, \citenamefont {Belov},\ and\
  \citenamefont {Kivshar}}]{Krasnok_11}%
  \BibitemOpen
  \bibfield  {author} {\bibinfo {author} {\bibfnamefont {A.~E.}\ \bibnamefont
  {Krasnok}}, \bibinfo {author} {\bibfnamefont {A.~E.}\ \bibnamefont
  {Miroshnichenko}}, \bibinfo {author} {\bibfnamefont {P.~A.}\ \bibnamefont
  {Belov}}, \ and\ \bibinfo {author} {\bibfnamefont {Y.~S.}\ \bibnamefont
  {Kivshar}},\ }\href@noop {} {\bibfield  {journal} {\bibinfo  {journal} {JETP
  Lett.}\ }\textbf {\bibinfo {volume} {94}},\ \bibinfo {pages} {635} (\bibinfo
  {year} {2011})}\BibitemShut {NoStop}%
\bibitem [{\citenamefont {Filonov}\ \emph {et~al.}(2012)\citenamefont
  {Filonov}, \citenamefont {Krasnok}, \citenamefont {Slobozhanyuk},
  \citenamefont {Kapitanova}, \citenamefont {Nenasheva}, \citenamefont
  {Kivshar},\ and\ \citenamefont {Belov}}]{Krasnok_APL_12}%
  \BibitemOpen
  \bibfield  {author} {\bibinfo {author} {\bibfnamefont {D.~S.}\ \bibnamefont
  {Filonov}}, \bibinfo {author} {\bibfnamefont {A.~E.}\ \bibnamefont
  {Krasnok}}, \bibinfo {author} {\bibfnamefont {A.~P.}\ \bibnamefont
  {Slobozhanyuk}}, \bibinfo {author} {\bibfnamefont {P.~V.}\ \bibnamefont
  {Kapitanova}}, \bibinfo {author} {\bibfnamefont {E.~A.}\ \bibnamefont
  {Nenasheva}}, \bibinfo {author} {\bibfnamefont {Y.~S.}\ \bibnamefont
  {Kivshar}}, \ and\ \bibinfo {author} {\bibfnamefont {P.~A.}\ \bibnamefont
  {Belov}},\ }\href@noop {} {\bibfield  {journal} {\bibinfo  {journal} {Appl.
  Phys. Lett.}\ }\textbf {\bibinfo {volume} {100}},\ \bibinfo {pages} {201113}
  (\bibinfo {year} {2012})}\BibitemShut {NoStop}%
\bibitem [{\citenamefont {Krasnok}\ \emph {et~al.}(2012)\citenamefont
  {Krasnok}, \citenamefont {Miroshnichenko}, \citenamefont {Belov},\ and\
  \citenamefont {Kivshar}}]{KrasnokOE}%
  \BibitemOpen
  \bibfield  {author} {\bibinfo {author} {\bibfnamefont {A.~E.}\ \bibnamefont
  {Krasnok}}, \bibinfo {author} {\bibfnamefont {A.~E.}\ \bibnamefont
  {Miroshnichenko}}, \bibinfo {author} {\bibfnamefont {P.~A.}\ \bibnamefont
  {Belov}}, \ and\ \bibinfo {author} {\bibfnamefont {Y.~S.}\ \bibnamefont
  {Kivshar}},\ }\href@noop {} {\bibfield  {journal} {\bibinfo  {journal}
  {Optics Express}\ }\textbf {\bibinfo {volume} {20}},\ \bibinfo {pages}
  {20599} (\bibinfo {year} {2012})}\BibitemShut {NoStop}%
\bibitem [{\citenamefont {Evlyukhin}\ \emph {et~al.}(2012)\citenamefont
  {Evlyukhin}, \citenamefont {Novikov}, \citenamefont {Zywietz}, \citenamefont
  {Eriksen}, \citenamefont {Reinhardt}, \citenamefont {Bozhevolnyi},\ and\
  \citenamefont {Chichkov}}]{Evlyukhin}%
  \BibitemOpen
  \bibfield  {author} {\bibinfo {author} {\bibfnamefont {A.~B.}\ \bibnamefont
  {Evlyukhin}}, \bibinfo {author} {\bibfnamefont {S.~M.}\ \bibnamefont
  {Novikov}}, \bibinfo {author} {\bibfnamefont {U.}~\bibnamefont {Zywietz}},
  \bibinfo {author} {\bibfnamefont {R.~L.}\ \bibnamefont {Eriksen}}, \bibinfo
  {author} {\bibfnamefont {C.}~\bibnamefont {Reinhardt}}, \bibinfo {author}
  {\bibfnamefont {S.~I.}\ \bibnamefont {Bozhevolnyi}}, \ and\ \bibinfo {author}
  {\bibfnamefont {B.~N.}\ \bibnamefont {Chichkov}},\ }\href@noop {} {\bibfield
  {journal} {\bibinfo  {journal} {Nano Lett.}\ }\textbf {\bibinfo {volume}
  {12}},\ \bibinfo {pages} {3749} (\bibinfo {year} {2012})}\BibitemShut
  {NoStop}%
\bibitem [{\citenamefont {Kuznetsov}\ \emph {et~al.}(2012)\citenamefont
  {Kuznetsov}, \citenamefont {Miroshnichenko}, \citenamefont {Fu},
  \citenamefont {Zhang},\ and\ \citenamefont {Lukyanchuk}}]{Kuznetsov}%
  \BibitemOpen
  \bibfield  {author} {\bibinfo {author} {\bibfnamefont {A.~I.}\ \bibnamefont
  {Kuznetsov}}, \bibinfo {author} {\bibfnamefont {A.~E.}\ \bibnamefont
  {Miroshnichenko}}, \bibinfo {author} {\bibfnamefont {Y.~H.}\ \bibnamefont
  {Fu}}, \bibinfo {author} {\bibfnamefont {J.}~\bibnamefont {Zhang}}, \ and\
  \bibinfo {author} {\bibfnamefont {B.}~\bibnamefont {Lukyanchuk}},\
  }\href@noop {} {\bibfield  {journal} {\bibinfo  {journal} {Sci. Rep.}\
  }\textbf {\bibinfo {volume} {2}},\ \bibinfo {pages} {492} (\bibinfo {year}
  {2012})}\BibitemShut {NoStop}%
\bibitem [{\citenamefont {Rolly}\ \emph {et~al.}(2011)\citenamefont {Rolly},
  \citenamefont {Stout}, \citenamefont {Bidault},\ and\ \citenamefont
  {Bonod}}]{RollyOL}%
  \BibitemOpen
  \bibfield  {author} {\bibinfo {author} {\bibfnamefont {B.}~\bibnamefont
  {Rolly}}, \bibinfo {author} {\bibfnamefont {B.}~\bibnamefont {Stout}},
  \bibinfo {author} {\bibfnamefont {S.}~\bibnamefont {Bidault}}, \ and\
  \bibinfo {author} {\bibfnamefont {N.}~\bibnamefont {Bonod}},\ }\href@noop {}
  {\bibfield  {journal} {\bibinfo  {journal} {Opt. Lett.}\ }\textbf {\bibinfo
  {volume} {36}},\ \bibinfo {pages} {3368} (\bibinfo {year}
  {2011})}\BibitemShut {NoStop}%
\bibitem [{\citenamefont {Alu}\ and\ \citenamefont {Engheta}(2007)}]{Alu}%
  \BibitemOpen
  \bibfield  {author} {\bibinfo {author} {\bibfnamefont {A.}~\bibnamefont
  {Alu}}\ and\ \bibinfo {author} {\bibfnamefont {N.}~\bibnamefont {Engheta}},\
  }\href@noop {} {\bibfield  {journal} {\bibinfo  {journal} {IEEE Trans. on
  Antennas and Propagation}\ }\textbf {\bibinfo {volume} {55}},\ \bibinfo
  {pages} {3027} (\bibinfo {year} {2007})}\BibitemShut {NoStop}%
\bibitem [{\citenamefont {Palik}(1985)}]{Palik}%
  \BibitemOpen
  \bibfield  {author} {\bibinfo {author} {\bibfnamefont {E.}~\bibnamefont
  {Palik}},\ }\href@noop {} {\emph {\bibinfo {title} {Handbook of Optical
  Constant of Solids}}}\ (\bibinfo  {publisher} {San Diego, Academic},\
  \bibinfo {year} {1985})\BibitemShut {NoStop}%
\bibitem [{\citenamefont {Boriskina}\ \emph {et~al.}(2006)\citenamefont
  {Boriskina}, \citenamefont {Benson}, \citenamefont {Sewell},\ and\
  \citenamefont {Nosich}}]{Sveta1}%
  \BibitemOpen
  \bibfield  {author} {\bibinfo {author} {\bibfnamefont {S.}~\bibnamefont
  {Boriskina}}, \bibinfo {author} {\bibfnamefont {T.}~\bibnamefont {Benson}},
  \bibinfo {author} {\bibfnamefont {P.}~\bibnamefont {Sewell}}, \ and\ \bibinfo
  {author} {\bibfnamefont {A.}~\bibnamefont {Nosich}},\ }\href@noop {}
  {\bibfield  {journal} {\bibinfo  {journal} {IEEE J. Select. Topics Quantum
  Electron.}\ }\textbf {\bibinfo {volume} {12}},\ \bibinfo {pages} {1175}
  (\bibinfo {year} {2006})}\BibitemShut {NoStop}%
\bibitem [{\citenamefont {Wang}\ \emph {et~al.}(2010)\citenamefont {Wang},
  \citenamefont {Yan}, \citenamefont {Yu}, \citenamefont {Unterhinninghofen},
  \citenamefont {Wiersig}, \citenamefont {Pflugl}, \citenamefont {Diehl},
  \citenamefont {Edamura}, \citenamefont {Yamanishi}, \citenamefont {Kan},\
  and\ \citenamefont {Capasso}}]{Wang}%
  \BibitemOpen
  \bibfield  {author} {\bibinfo {author} {\bibfnamefont {Q.~J.}\ \bibnamefont
  {Wang}}, \bibinfo {author} {\bibfnamefont {C.}~\bibnamefont {Yan}}, \bibinfo
  {author} {\bibfnamefont {N.}~\bibnamefont {Yu}}, \bibinfo {author}
  {\bibfnamefont {J.}~\bibnamefont {Unterhinninghofen}}, \bibinfo {author}
  {\bibfnamefont {J.}~\bibnamefont {Wiersig}}, \bibinfo {author} {\bibfnamefont
  {C.}~\bibnamefont {Pflugl}}, \bibinfo {author} {\bibfnamefont
  {L.}~\bibnamefont {Diehl}}, \bibinfo {author} {\bibfnamefont
  {T.}~\bibnamefont {Edamura}}, \bibinfo {author} {\bibfnamefont
  {M.}~\bibnamefont {Yamanishi}}, \bibinfo {author} {\bibfnamefont
  {H.}~\bibnamefont {Kan}}, \ and\ \bibinfo {author} {\bibfnamefont
  {F.}~\bibnamefont {Capasso}},\ }\href@noop {} {\bibfield  {journal} {\bibinfo
   {journal} {PNAS}\ }\textbf {\bibinfo {volume} {107}},\ \bibinfo {pages}
  {22407} (\bibinfo {year} {2010})}\BibitemShut {NoStop}%
\bibitem [{\citenamefont {Scully}(2010)}]{Scully}%
  \BibitemOpen
  \bibfield  {author} {\bibinfo {author} {\bibfnamefont {M.~O.}\ \bibnamefont
  {Scully}},\ }\href@noop {} {\bibfield  {journal} {\bibinfo  {journal} {PNAS}\
  }\textbf {\bibinfo {volume} {107}},\ \bibinfo {pages} {22367} (\bibinfo
  {year} {2010})}\BibitemShut {NoStop}%
\bibitem [{\citenamefont {Jackson}(1998)}]{Jackson}%
  \BibitemOpen
  \bibfield  {author} {\bibinfo {author} {\bibfnamefont {J.}~\bibnamefont
  {Jackson}},\ }\href@noop {} {\emph {\bibinfo {title} {Classical
  Electrodynamics}}}\ (\bibinfo  {publisher} {New York : Wiley},\ \bibinfo
  {year} {1998})\BibitemShut {NoStop}%
\end{thebibliography}

%

\end{document}